\documentclass{PoS}
\usepackage{lineno}

\title{Searching for Very High Energy Emission from Pulsars Using the High Altitude Water Cherenkov (HAWC) Observatory}

\ShortTitle{Search for VHE Emission from Pulsars with HAWC}

\author{\speaker{Alvarez Ochoa, C.}$^a$, Saz Parkinson, P.~M.$^{b,c}$, Belfiore, A.$^d$, Carrami\~ nana, A.$^e$, Rivi\`{e}re, C.$^f$, and Moreno, E.$^g$ for the HAWC Collaboration$^h$ \\
  \llap{$^a$}Facultad de Ciencias en F\'{i}sica y Matem\'{a}ticas, Universidad Aut\'{o}noma de Chiapas, M\'{e}xico \\
  \llap{$^b$}Department of Physics, University of Hong Kong, Pokfulam, Hong Kong \\
  \llap{$^c$}Santa Cruz Institute for Particle Physics, University of California, Santa Cruz, CA 95064, USA \\
  \llap{$^d$}INAF -- IASF Milano, Via E. Bassini 15, I-20133 Milano, Italy\\
  \llap{$^e$}Instituto Nacional de Astrof\'{i}sica, \'{O}ptica y Electr\'{o}nica, Tonantzintla, Puebla, Mexico \\
  \llap{$^f$}Department of Physics, University of Maryland, College Park, MD 20742-4111, USA \\
  \llap{$^g$}Facultad de Ciencias F\'{i}sico Matem\'{a}ticas, Benem\'{e}rita Universidad Aut\'{o}noma de Puebla, Ciudad Universitaria, Colonia San Manuel Puebla, M\'{e}xico \\
  \llap{$^h$}For a complete author list, see the special section of these proceedings.
        Email: \email{cesar.alvarez@unach.mx}, \email{pablosp@hku.hk}, \email{mario@piffio.org}, \email{alberto@inaoep.mx}, \email{riviere@udmgrb.umd.edu}, \email{emoreno@fcfm.buap.mx}}

\abstract{There are currently over 160 known gamma-ray pulsars. While most of them are detected only from space, at least two are now seen also from the ground. MAGIC and VERITAS have measured the gamma ray pulsed emission of the Crab pulsar up to hundreds of GeV and more recently MAGIC has reported emission at $\sim$2~TeV. Furthermore, in the Southern Hemisphere, H.E.S.S. has detected the Vela pulsar above 30~GeV. In addition, non-pulsed TeV emission coincident with pulsars has been detected by many groups, including the Milagro Collaboration. These GeV--TeV observations open the possibility of searching for very-high-energy (VHE, > 100~GeV) pulsations from gamma-ray pulsars in the HAWC field of view. 
}

\FullConference{The 34th International Cosmic Ray Conference,\\
		30 July- 6 August, 2015\\
		The Hague, The Netherlands}

\begin{document}

\section{Introduction}

Pulsars are rapidly-spinning, highly-magnetized neutron stars.  The pulsed emission from pulsars has been detected from the radio to gamma rays regime. Since its launch in 2008, the {\it Fermi} Large Area Telescope (LAT) has detected more than 160 gamma-ray pulsars reported in a regularly updated public catalog\footnote{{\tt https://confluence.slac.stanford.edu/x/5Jl6Bg}}. Fermi-LAT 4 year point source catalog (3FGL) has 147 gamma-ray pulsars detected at energies above 100~MeV~\cite{Acero15}, an increase of over an order of magnitude compared to the number of gamma-ray pulsars detected by previous-generation experiments. Fermi-LAT high energy point source catalog (1FHL) has 28 pulsars detected above 10 GeV, and 13 pulsars showed emission above 25~GeV~\cite{Ackermann13}. Beyond these energies it becomes hard to study pulsars with the LAT (though not impossible, see for example~\cite{Leung14,McCann15}), given the few photons detected (a consequence of both the steeply falling spectra and the relatively small effective area of the instrument). Thus, the much larger effective areas of ground-based (Cherenkov) gamma-ray instruments make them better suited for studies of pulsars at very high energies, while the critical $\sim$50--100~GeV energy range, can be considered a critical overlap range, where detections might be possible both from the ground and from space. Pulsed gamma-ray emission above 25~GeV was first detected from the Crab pulsar by the MAGIC telescope~\cite{Aliu08}. This was followed up with the detection of pulsed emission above 100~GeV by the VERITAS Collaboration~\cite{Aliu11}, an unexpected result which presented serious challenges to most conventional pulsar emission models, leading to a flurry of theoretical activity attempting to explain such emission~\cite{Du12,Lyutikov12,Petri12,Bednarek12,Aharonian12}. The detection of pulsars above 100~GeV also makes them interesting sources for studies of Lorentz Invariance Violation~\cite{Otte11,Nellen15}. 

Recently, the MAGIC Collaboration reported a detection of pulsed emission from the Crab at $\sim$2~TeV \cite{Zanin14}. Whether the Crab pulsar is unique in this respect is something that remains to be seen. Unfortunately, despite their good sensitivities above 100~GeV, Atmospheric Cherenkov Telescopes (ACTs) like MAGIC, VERITAS, and H.E.S.S. have relatively small fields of view and duty cycles, which limits the number of pulsars that can be explored. The recent detection of pulsed gamma-ray emission from the Vela pulsar above 50~GeV by H.E.S.S.\footnote{{\tt http://www.mpg.de/8287998/velar-pulsar}} demonstrates the rapid progress in the field, but also highlights the difficulties in finding another pulsar, like the Crab, to test the proposed emission models under different parameters. 

\section{HAWC}

The High Altitude Water Cherenkov gamma-ray observatory (HAWC)~\cite{Stefan14}, inaugurated in March 2015, is located at 19$^\circ$N in latitude and 4,100~m above sea level, on the volcano Sierra Negra in the state of Puebla, Mexico. HAWC consists of 300 tanks, 7.3~m in diameter and 4.5~m deep, spread out over an area of 22,000~m$^2$, each containing $\sim$200,000~L of ultra-pure water in a water-tight bladder. The bottom of each tank is instrumented with four upward facing photomultiplier tubes (PMTs) sensitive to ultraviolet light-tight. One large (10''), high-quantum-efficiency, PMT is located at the center, with three smaller (8'') PMTs reused from the Milagro experiment ~\cite{Abdo07}, placed in an equilateral triangle around the center. HAWC has an instantaneous field of view of 2~sr with a duty cycle $>$ 95$\%$. Thanks to its modular design, it was possible to operate the partially built detector during deployment, before the completion of the final array. For example, the partial array referred as HAWC-111, operated between August 2, 2013 to July 8, 2014 with a air shower trigger rate of 15~kHz~\cite{Zhou15}, resulting in a $\sim24\sigma$ detection of the Crab Nebula over this period~\cite{Salesa15}. For an overview of the results, see for instance~\cite{Pretz15}.  
 
\section{Pulsar Observation with HAWC}

Thanks to its wide field of view and high duty cycle, HAWC is able to observe more than 100 of known gamma-ray pulsars for an extended period of time. One scientific goal of HAWC is to search for pulsed very-high-energy gamma-ray emission from these pulsars. One challenge of HAWC for detecting pulsed emission from pulsars is the higher contamination of cosmic-ray events in the data set. HAWC detects air showers in a rate of 15~kHz. However, only less than 0.1\% of them are gamma rays. 
Therefore in order to detect faint sources like pulsars, it is very important to have a powerful method for discriminating gamma-ray showers from hadronic showers. Multiple efforts are underway within the collaboration to improve the gamma/hadron rejection~\cite{Capistran15,Hampel15}. Another challenge is to estimate the background contamination from the associated pulsar wind nebulae. In particularly young pulsars like Crab pulsar have an associated pulsar wind nebula that is bright in very-high-energy gamma-ray regime~\cite{Bartoli15}. 

For the high-energy pulsars with known ephemeris, on and off-phases can be used to estimate the background contamination from both hadronic showers and the associated nebula. For example VERITAS found~\cite{Aliu11} that all of Crab pulsar gamma ray events arrive in two peaks that cover only $\sim 1/15^{\textrm{th}}$ of the phase, which is called on-phase. There are no gamma rays coming from the pulsar in the remaining $\sim 14/15^{\textrm{th}}$ of the phase, which is called off-phase. Since there are no gamma rays from the pulsar in the off-phase that part can be used to estimate the backgrounds. This method will be able to drastically improve the sensitivity of the HAWC for pulsars. However, in order to apply this method we should make sure that HAWC has accurate timing measurements.

\subsection{Timing Validation}

 In order to demonstrate that the HAWC DAQ and analysis is able to perform such a search, 
an optical system is under development. A 12" MEAD LX200 telescope will be used to look at the Crab in the optical domain. Light will be collected by a small PMT and will be sent - after shaping - to the regular 
HAWC DAQ, where signals will be timestamped with the same timing system as in the standard analysis. A pulsation search will then be performed, by assigning a phase to each photon using the pulsar ephemeris from the Jodrell Bank Observatory\footnote{{\tt http://www.jb.man.ac.uk/pulsar/crab.html}} and the TEMPO-2 framework~\cite{Hobbs06}. This will validate the timing system of HAWC and the phase calculation; thus the capacity to perform pulsation searches with HAWC will be improved, whether to observe pulsation or to set limits.
For the actual searches using gamma-ray data, same pipeline will be used for gamma-ray pulsar searches.

\section{Conclusions}

  
The recent inauguration of the HAWC observatory, a wide-field high-duty-cycle gamma-ray instrument sensitive in the 100~GeV -- 100~TeV energy range, opens up the possibility of carrying out a search for VHE emission from a large number of pulsars in its field of view.

\section*{Acknowledgments}
\footnotesize{
We acknowledge the support from: the US National Science Foundation (NSF);
the US Department of Energy Office of High-Energy Physics;
the Laboratory Directed Research and Development (LDRD) program of
Los Alamos National Laboratory; Consejo Nacional de Ciencia y Tecnolog\'{\i}a (CONACyT),
Mexico (grants 260378, 55155, 105666, 122331, 132197, 167281, 167733,  207912);
Red de F\'{\i}sica de Altas Energ\'{\i}as, Mexico;
DGAPA-UNAM (grants IG100414-3, IN108713,  IN121309, IN115409, IN111315);
VIEP-BUAP (grant 161-EXC-2011);
the University of Wisconsin Alumni Research Foundation;
the Institute of Geophysics, Planetary Physics, and Signatures at Los Alamos National Laboratory;
the Luc Binette Foundation UNAM Postdoctoral Fellowship program. We are grateful to The University of 
California Institute for Mexico and the United States (UC MEXUS) and ``El Consejo Nacional de Ciencia y Tecnolog\'{i}a
(CONACYT)'' for their support of this research, in the form of a Collaborative Grant.
}

\def \apj {ApJ}
\def \aap {A\&A}
\def \apjs {ApJS}
\def \mnras {MNRAS}
\def \apjl {ApJL}

\bibliographystyle{JHEP}
\bibliography{icrc2015-1369}

\end{document}